\documentclass[notitlepage,letterpaper,aps,twocolumn,amsmath,amsfonts,nofootinbib,preprintnumbers,superscriptaddress,eqsecnum,secnumarabic]{revtex4-1}
\pdfoutput=1
\usepackage{amssymb,amsmath,latexsym,mathrsfs}
\usepackage{bm}
\usepackage{url}
\usepackage{epsfig,graphicx,verbatim,xspace,multirow,mathtools}
\usepackage{array}
\usepackage{enumitem}
\usepackage{graphicx,slashed}
\usepackage[usenames,dvipsnames]{color}
\usepackage[normalem]{ulem}
\usepackage[breaklinks,colorlinks,urlcolor=blue,citecolor=blue,linkcolor=blue]{hyperref}
\usepackage{grffile}
\usepackage[toc,page]{appendix}

\begin{document}

\title{Removing Astrophysics in 21 cm maps with Neural Networks}

\author{Pablo Villanueva-Domingo}
\email{pablo.villanueva@ific.uv.es}
\affiliation{Instituto de F\'isica Corpuscular (IFIC),
  CSIC-Universitat de Valencia,\\
Apartado de Correos 22085,  E-46071, Spain}
\author{Francisco Villaescusa-Navarro}
\email{fvillaescusa@princeton.edu}
\affiliation{Department of Astrophysical Sciences, Princeton University, Peyton Hall, Princeton NJ 08544, USA}
\affiliation{Center for Computational Astrophysics, Flatiron Institute, 162 5th Avenue, 10010, New York, NY, USA}

\begin{abstract}
Measuring temperature fluctuations in the 21 cm signal from the Epoch of Reionization and the Cosmic Dawn is one of the most promising ways to study the Universe at high redshifts. Unfortunately, the 21 cm signal is affected by both cosmology and astrophysics processes in a non-trivial manner. We run a suite of 1,000 numerical simulations with different values of the main astrophysical parameters. From these simulations we produce tens of thousands of 21 cm maps at redshifts $10\leq z\leq 20$. We train a convolutional neural network to remove the effects of astrophysics from the 21 cm maps, and output maps of the underlying matter field. We show that our model is able to generate 2D matter fields that not only resemble the true ones visually, but whose statistical properties agree with the true ones within a few percent down to scales $\simeq 2 {\rm Mpc}^{-1}$. We demonstrate that our neural network retains astrophysical information, that can be used to constrain the value of the astrophysical parameters. Finally, we use saliency maps to try to understand which features of the 21 cm maps the network is using in order to determine the value of the astrophysical parameters.

\medskip

\textbf{Key words:} cosmology: dark ages, reionization, first stars - large-scale structure of Universe - dark matter - galaxies: intergalactic medium - high redshift - methods: statistical

\end{abstract}

\maketitle

\section{Introduction}

Measuring the cosmological 21 cm signal represents one of the most promising probes to gain insight into the evolution of the Intergalactic Medium (IGM), as well as an observational challenge. On the one hand, detecting the global average temperature is the main goal of several current and future experiments, such as \texttt{EDGES} (Experiment to Detect the Global EoR Signature) \citep{Bowman:2018yin}, \texttt{LEDA} (Large Aperture Experiment to Detect the Dark Ages)~\citep{Greenhill:2012mn} and \texttt{DAPPER} (Dark Ages Polarimeter PathfindER)~\citep{2019AAS...23421202B}. The observation of an absorption profile located at a redshift of $z \sim 17$ has been recently claimed by the \texttt{EDGES} experiment, with an amplitude which is twice the maximum predicted within the context of the $\Lambda$CDM model~\citep{Bowman:2018yin}. This has motivated a large number of studies in the literature (see, e.g., \cite{Munoz:2018pzp, Barkana:2018lgd, Fialkov:2018xre, Mirocha:2018cih, Pospelov:2018kdh, Witte:2018itc, Lopez-Honorez:2018ipk, Fialkov:2019vnb, Ewall-Wice:2019may}). 

Meanwhile, other experiments are devoted to observe the fluctuations of the differential brightness temperature, which allows a better foreground removal and contain additional information beyond the one from the global signal. Among them, there are some examples which have already collected data, such as \texttt{GMRT} (Giant Metrewave Radio Telescope), \texttt{LOFAR} (LOw Frequency ARray) \citep{van_Haarlem_2013}, \texttt{MWA} (Murchison Widefield Array) \citep{Bowman_2013} and \texttt{PAPER} (Precision Array for Probing the Epoch of Reionization) \citep{Ali:2015uua}, while others like \texttt{HERA} (Hydrogen Epoch of Reionization Array) \citep{DeBoer_2017} or \texttt{SKA} (Square Kilometer Array) \citep{Mellema:2012ht} are expected to be completed within this decade.

The aim of these experiments is to shed light on the poorly known astrophysics that shaped the reionization of the Universe. The amplitude, time dependence and spatial distribution of the 21 cm signal arises not only from astrophysics, but is also sensitive to cosmology. Therefore, 21 cm maps can be used to improve our knowledge not only of astrophysics, but also about cosmology. Unfortunately, the signatures left by astrophysics and cosmology on the 21 cm signal are coupled in a non-trivial manner: the spatial distribution of matter, with its peaks (halos) positions and sizes (masses) is mainly driven by cosmology. In those halos, gas is able to cool down and form stars and black holes. The radiation from those objects ionize neutral hydrogen, creating bubbles that can expand tens of Megaparsecs. The 21 cm signal is sensitive to all these astrophysical processes. 

Ideally, we would like to constrain both cosmology and astrophysics from 21 cm observations. Lots of work has been done in extracting this information from either power spectrum or other summary statistics \citep{Mao_2008, Mesinger_2013, Greig_2015, Liu_2016, Cohen_2018, Park:2018ljd, Majumdar_2018, Watkinson_2018, Sitwell_2014, Mu_oz_2019_1, Mu_oz_2019_2}. In this work we take a different route and try to \textit{undo} the effects of astrophysics on 21 cm maps. By doing that, we will obtain a map with the spatial distribution of matter in the high-redshift Universe. Those maps will be an invaluable source of information. First, they will represent a picture of the Universe at high-redshift, allowing us to cross-check the validity of the $\Lambda$CDM model. Second, since those maps will be obtained at high-redshift, the power spectrum will be able to retrieve all cosmological information down to very small scales. Third, those maps can be used for cross-correlation studies with different tracers, e.g. high-redshift galaxies, to learn about the halo-galaxy connection. Fourth, those maps will allow us better reconstruct the effect of astrophysics, and its correlation with the matter field.

In this paper, we make use of machine learning techniques to \textit{undo} the effects of astrophysics and produce 2D matter density fields in real-space from 21 cm maps in redshift-space. In particular, we made use of deep convolutional neural networks. Recently, artificial neural networks have been broadly applied to cosmology, and specifically, to study the 21 cm signal. Examples of this are extracting the global signal from different foreground models \citep{Choudhury:2019vat}, generating accurate HI fields and 21 cm maps by means of Generative Adversarial Networks \citep{Zamudio-Fernandez:2019lxp,List:2020qdz,feder2020nonlinear}, and emulating reionization simulations to extract information about the underlying astrophysical processes, by means of the 21 cm power spectrum  \citep{Schmit:2017pho,Kern:2017ccn,Shimabukuro:2017jdh} or by studying the tomography of the HI fields, directly from 21 cm maps \citep{Gillet:2018fgb,Hassan:2018uhw,Hassan:2019cal,kwon2020deeplearning}\footnote{For a comprehensive list of applications of neural networks (and machine learning in general) to cosmology, see \url{https://github.com/georgestein/ml-in-cosmology}}.

Our goal is to train a neural network to output the projected dark matter field in real-space, from a 21 cm map in redshift-space generated with given astrophysics. Therefore, we want to undo both the effects of astrophysics but also of redshift-space distortions. By feeding the network with 21 cm maps produced by different values of the astrophysical parameters, and different realizations (due to cosmic variance), we are forcing the model to learn to undo the effects of astrophysics, within their range of variation. We focus on the redshift range between 20 and 10, where astrophysical heating and Ly$\alpha$ coupling become important. For lower redshifts, however, the map between density fluctuations and brigtness temperature becomes more complex, and the results of the network worsen, as we shall further show.

We have run 1,000 numerical simulations with different values of three of the most relevant astrophysical parameters, making use of the publicly available code \texttt{21cmFAST} \citep{Mesinger:2010ne,Park:2018ljd}. From these simulations we then extract 120 2D slices per simulation of 21 cm and matter maps that we use as training, validation and test sets for our neural network.

The paper is organized as follows. In Sec. \ref{sec:21cm}, we briefly describe the fundamentals of the 21 cm signal and the relevant astrophysical parameters which can leave an imprint on it. We discuss the methods we use in Sec. \ref{sec:methods}. We then present the main results of this work in Sec. \ref{sec:results}. Finally, we draw the main conclusions in Sec. \ref{sec:conclusions}.

\section{The 21 cm signal}
\label{sec:21cm}

\subsection{Fundamentals of the 21 cm signal}

In this work we have made use of the publicly available code \texttt{21cmFAST} \citep{Mesinger:2010ne,Park:2018ljd} to compute the matter density and 21 cm maps. We start by reviewing some basic aspects of this redshifted line which are implemented in the code (see, e.g., Refs.~\cite{10.1088/2514-3433/ab4a73,Madau:1996cs, Furlanetto:2006jb, Pritchard:2011xb, Furlanetto:2015apc} for comprehensive reviews). This line arises from the splin-flip transition in the hyperfine structure of the hydrogen atom. The intensity of the line depends on the ratio between the excited and ground states of the $1s$ level of neutral hydrogen. This ratio is conventionally characterized by the so-called spin temperature $T_S$, and is formally defined through the next equation:
\begin{equation}
\frac{n_1}{n_0} = 3 \, e^{- T_0/ T_S} ~,
\end{equation}
where $n_0$ and $n_1$ are the number density of neutral hydrogen atoms in the ground (singlet) and excited (triplet) states, $T_0 = h \nu_0/k_B = 0.068$ K is the energy splitting of the hyperfine transition, while the factor of 3 comes from the degeneracy of the triplet excited state.

Three main processes can modify the population of the excited and ground levels, and determine therefore the spin temperature: $\emph{(i)}$ absorption and stimulated emission induced by a background radiation field, namely the Cosmic Microwave Background (CMB); $\emph{(ii)}$ collisions of neutral hydrogen atoms with other hydrogen atoms, free protons, or free electrons; and $\emph{(iii)}$ indirect splin-flip transitions produced by scattering with Lyman-$\alpha$ photons, the so-called Wouthuysen-Field effect \citep{Wouthuysen:1952,Field:1958}. The spin temperature can be expressed as a weighted sum over the temperatures which characterize the efficiency of each of these processes:
\begin{equation}\label{eq:spinT}
T_S = \frac{T_{\rm CMB} + y_k \, T_k + y_\alpha \, T_\alpha}{1 + y_k + y_\alpha} ~,
\end{equation}
where $T_{\rm CMB}$, $T_k$, and $T_\alpha$ are the temperature of the CMB, the kinetic temperature of the gas, and the color temperature, respectively. The latter is associated with the intensity of the Lyman-$\alpha$ emission, and for most cases of interest, $T_\alpha \simeq T_k$ (although 21cmFAST computes it properly by an iterative procedure). The strength of the different processes are given by the coupling coefficients, $y_k$ for collisions, and $y_\alpha$ for Ly$\alpha$ scatttering (see, e.g., \cite{Furlanetto:2006jb, Pritchard:2011xb} for more details).

It is customary to express the intensity of the 21 cm line relative to the CMB in terms of the differential brightness temperature $\delta T_b$. Taking into account absorption and emission of 21 cm photons in the expanding IGM, it is possible to write the solution of the radiative transfer equation as
\begin{equation}
\delta T_b = \frac{T_S - T_{\rm CMB}}{1 + z} (1 - e^{-\tau_{\nu_0}}) ~,
\label{eq:Tb}
\end{equation}
where $\tau_{\nu_0}$ is the optical depth of the 21 cm line. Since the value of the optical depth, at the redshifts of interest, is always small, we can safely expand the exponential term and express the differential brightness temperature as \citep{Furlanetto:2006jb,Pritchard:2011xb}
\begin{align}
\delta T_b & \simeq 27 \, x_\textrm{HI} \, (1 + \delta) \left( 1 - \frac{T_\textrm{CMB}}{T_S}\right) \nonumber \\ & \times \left( \frac{1}{1+H^{-1} \partial v_r / \partial r} \right) \, \left( \frac{1+z}{10}\right)^{1/2} \nonumber \\ & \times \left(\frac{0.15}{\Omega_{\rm m} h^2} \right)^{1/2} \left( \frac{\Omega_{\rm b} h^2}{0.023}\right)\,\textrm{mK} ~,
\label{eq:Tbdev}
\end{align}
where $\delta=\rho/\bar{\rho}-1$ is the matter overdensity, $H$ is the Hubble parameter, $\partial v_r / \partial r$ is the velocity gradient along the line of sight, and $\Omega_{\rm m}$ and $\Omega_{\rm b}$ are the matter and baryon abundances of the Universe at $z=0$. We see in Eq. \ref{eq:Tbdev} that $\delta T_b$ depends directly on $\delta$, providing an indirect measurement of the underlying matter density field. Nevertheless, it also depends on several other quantities related with the ionization and thermal state of the hydrogen, such as the neutral fraction $x_\textrm{HI}$, the temperature of the gas and the Lyman-$\alpha$ background field, through the spin temperature $T_S$. Therefore, the link between the 21 cm field $\delta T_b$ and the density field $\delta$ is not trivial.

We emphasize that the above equation holds at every spatial location, $\vec{x}$. What it is observed is $\delta T_b(\vec{x})$, and without knowing quantities like $x_{\rm HI}(\vec{x})$ and $T_S(\vec{x})$, it is not possible to determine the value of $\delta(\vec{x})$. Fortunately, there are non-trivial spatial correlations between those fields and the underlying matter field. A neural network may be able to identify these correlations and use them to infer the value of $\delta(\vec{x})$ from $\delta T_b(\vec{x})$. This is the purpose of this work.

\subsection{Astrophysical parameters}
\label{sec:astro}

\begin{table*}[t]
	\setlength\extrarowheight{5pt}
	\begin{center}
	\renewcommand{\arraystretch}{0.8}
\resizebox{1.0\textwidth}{!}{
		\begin{tabular}{|c||c|c|c|}
			\hline
			Parameter & Range & Units & Description \\ 
			\hline\hline
			$L_{X \leq 2 {\rm keV}}/{\rm SFR}$ & $10^{40}-10^{41}$ & $\rm erg \, s^{-1} \, M_\odot^{-1} \, yr$ & Luminosity of X-rays per star formation rate \\
			\hline 
			$M_{\rm min}$ & $10^{8} - 10^{9}$ & $M_\odot$ & $\;$ Minimum mass of a halo to host star formation $\;$ \\
			\hline
			$N_{\gamma/b_\star}$ & $10^3 - 10^4$ & - & Number of ionizing photons per baryon \\
			\hline
		\end{tabular}}
	\end{center}
	\caption{\label{table:astro_params} We run 1,000 numerical simulations using \texttt{21cmFAST} varying the value of the above astrophysical parameters in the range indicated in the second column. The parameter values are arranged in a latin-hypercube.}
\end{table*}

Now we turn our attention to the relevant astrophysical processes which can leave an imprint on the thermal history of the Universe, and consequently, affect the amplitude and spatial distribution of the 21 cm signal. Here we assume a minimal model with three free parameters. We refer the reader to \cite{Mesinger:2010ne,Park:2018ljd} for more details on the astrophysical parameterization.

Firstly, we assume that only the most massive halos are able to cool down the gas to trigger star formation. Therefore, low-mass halos do not form stars and do not contribute to the X-ray and UV radiation. Following the \texttt{21cmFAST} parameterization \citep{Mesinger:2010ne,Park:2018ljd}, we take an exponential cutoff for the mass integration $\exp(-M_{\rm min}/M_h)$, with $M_h$ the mass of the halo and $M_{\rm min}$ the threshold mass, assumed to be redshift independent. This minimum mass could be related with a virial temperature of the halo of $\sim 10^4$ K, where the threshold of atomic cooling lies, which corresponds to $\sim 10^8~M_\odot$. Reasonable values for $M_{\rm min}$ lie within $10^8\lesssim M_{\rm min}/M_\odot \lesssim 10^{9}$.

One of the most important processes that affects the 21 cm signal is the X-ray emission from astrophysical sources, such as High Mass X-ray Binaries (HMXB), which are able to heat up the gas of the IGM. Since the specific properties of the heating sources at high redshift are not well known yet, we assume a power-law profile for the X-ray emissivity, with $L_X^0$ being the normalization and $\alpha_X$ the spectral index. We fix $\alpha_X = 1$, consistent with HMXB \footnote{We do not expect this choice to have a strong impact on the results, given that several degeneracies between the astrophysical parameters are present, and thus small changes in the spectral index may be countered by variations in, e.g., the luminosity.}. Notice that not all photons will affect the IGM. On the one hand, only X-rays with energy $\lesssim 2$keV are absorbed in the IGM, as photons with energies above this threshold have mean free paths larger than the Hubble length. On the other hand, low energy photons below some threshold $E_0$ (taken as $0.5$ keV) are absorbed locally in the galaxy and are not able to escape to the IGM. The relevant luminosity which determines the heating of the IGM is therefore given by the integration of the emissivity over the range mentioned above:
\begin{equation}
 \frac{L_{X \leq 2 \, {\rm keV}}}{{\rm SFR}} = \int_{E_0}^{2 \, {\rm keV}} \, dE \, \frac{L_X^0}{{\rm SFR}} \,  E^{-\alpha_X} \, ,
\end{equation}
where ${\rm SFR}$ is the star formation rate. Therefore, we employ $L_{X \leq 2 {\rm keV}}/{\rm SFR}$ as a free parameter. This is a rather important quantity since it determines how quickly and abruptly the gas is heated, affecting the amplitude and width of the 21 cm absorption profile. We let it vary in a physical relevant range between $10^{40}$ and $10^{41} \, {\rm ergs \, s^{-1}\, M_\odot^{-1} \, yr}$.

Lastly, we consider the modelling of the UV radiation, which is important for two reasons. First of all, UV photons redshifting to Lyman resonances in the high-z Universe can cascade to the Ly$\alpha$ line and couple the spin temperature to the kinetic temperature of the gas (the so-called Wouthuysen-Field effect), driving the 21 cm signal to an absorption period. Secondly, when the emission of UV photons is efficient enough, the HII regions around galaxies can grow and merge, ionizing the full IGM, which is known as Epoch of Reionization. Since the 21 cm signal is proportional to the fraction of neutral hydrogen, it is very sensitive to the Reionization Epoch and therefore to the number of ionizing photons. We account for the normalization of the UV luminosity by means of the number of UV photons per stellar baryon $N_{\gamma/b_\star}$, which we consider can range from $\sim 10^3 - 10^4$. We follow the default \texttt{21cmFAST} parameterization, where the spectrum normalization between the Ly$\alpha$ and Lyman limit is proportional to the value of $N_{\gamma/b_\star}$, assuming Population II stars. Note that other choices of astrophysical parameters would also be meaningful, such as the fraction of baryons into stars, assumed here to be $f_* = 0.05$. Given we focus on $z \gtrsim 10$ and the degenerations between some astrophysical parameters, the above choice seems appropriate. Furthermore, it presents the advantage of being easy to interpret regarding their impact on the 21 cm signal and IGM history.

Table \ref{table:astro_params} summarizes the three free astrophysical parameters and their ranges of variation considered. Notice that for more realistic scenarios we may consider additional free parameters \citep{Greig:2015qca,Park:2018ljd}, but for the sake of simplicity here we restrict to this minimal choice. It is convenient to note that adding more parameters may worsen the performance of the network, since neural networks usually generalize badly. There is no guarantee to maintain the effectiveness to predict the density fields in that case. Hence, this has to be regarded as a proof of concept, rather than a conclusive work.

\section{Methods}

\begin{figure*}[th]
    \centering
    \includegraphics[width=\textwidth]{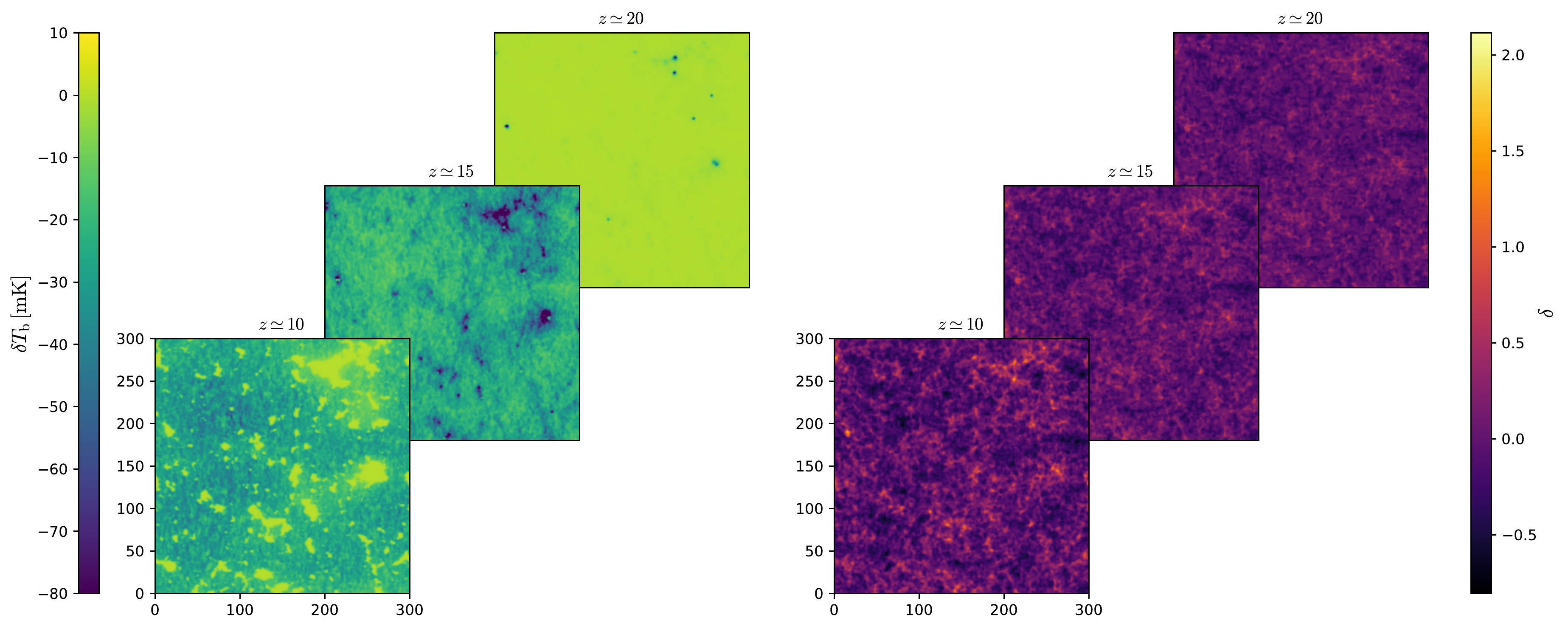}
    \caption{2D slices of the differential brightness temperature $\delta T_b$ in redshift-space (left) and the matter density field in real-space $\delta$ (right) at redshifts 10, 15, and 20. The purpose of this work is to train neural networks that find the mapping between the 21 cm and dark matter maps.}
    \label{fig:slices_redshift}
\end{figure*}

\begin{figure*}[th]
    \centering
    \includegraphics[width=\textwidth]{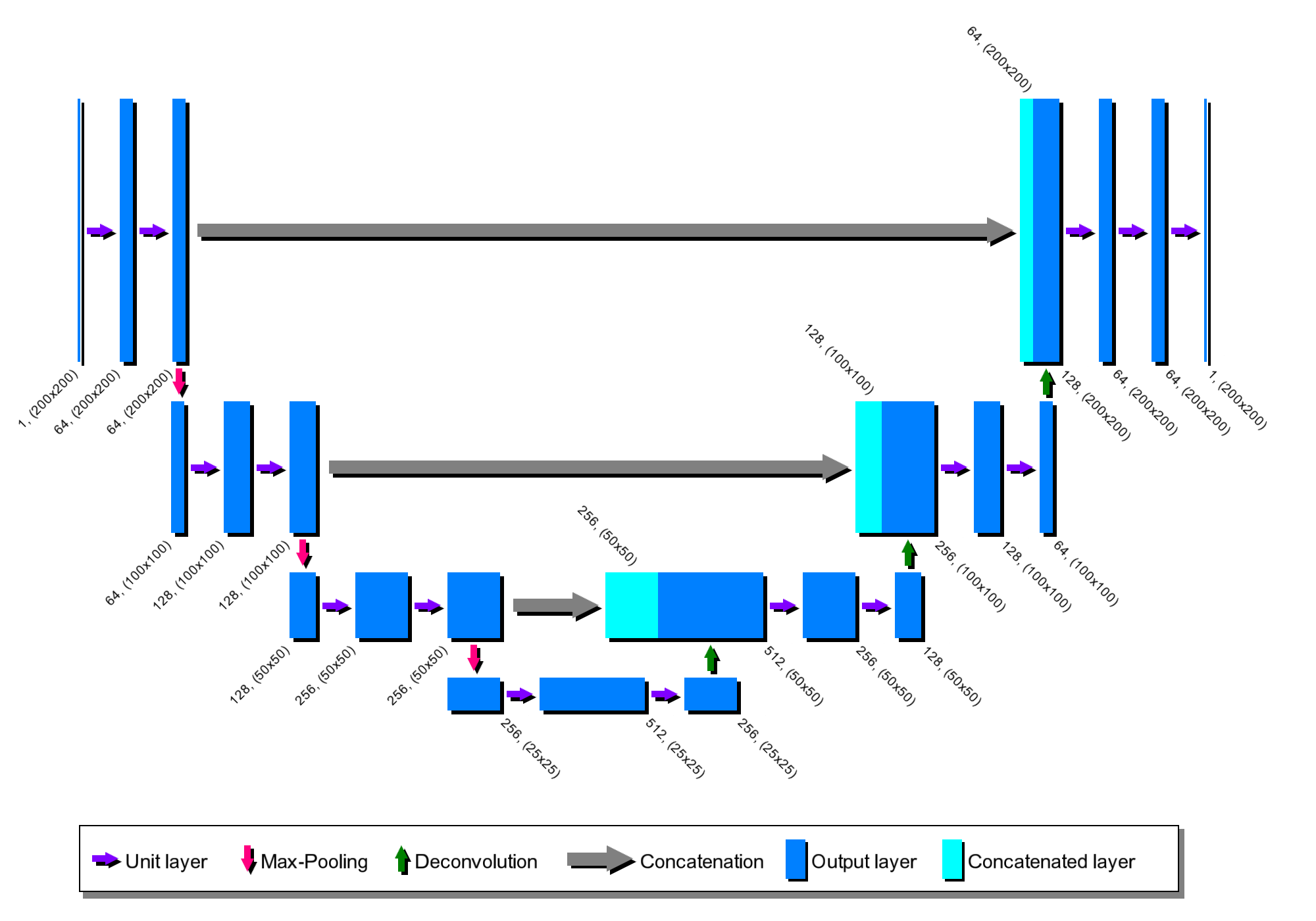}
    \caption{Scheme of the architecture employed in this work. The numbers associated to each output layer represent the number of channels (depth) and the spatial dimensions.}
    \label{fig:architecture}
\end{figure*}

\label{sec:methods}

In this section, we describe the numerical simulations run, the  architecture of the neural network used, and some caveats of the training process.

\subsection{Simulations}

We use the publicly available code \texttt{21cmFAST}\footnote{\url{https://github.com/andreimesinger/21cmFAST}} \citep{Mesinger:2010ne,Park:2018ljd} to generate the 21 cm and matter density fields at different redshifts. From an initial linear Gaussian field, this software employs several semi-analytical approximations to evolve the density field, produce a map of collapsed regions and compute the 21 cm brightness temperature. The 21 cm cubes are generated taking into account redshift-space distortions. The 3D box of the simulations contains $200^3$ voxels, with a physical length of $300$ Mpc, having therefore a spatial resolution of $1.5$ Mpc. To generate the density field, it has been employed second order lagrangian perturbation theory (2LPT) \citep{1998MNRAS.299.1097S}.

We run 1,000 simulations of 3D boxes (\textit{coeval} cubes in the \texttt{21cmFAST} terminology) with different initial random seeds\footnote{We have repeated the whole analysis of this paper using simulations with the same initial random seed. In this case, the network is able to reconstruct the dark matter field with a much higher accuracy. Notice that in this case, the network will have to predict always the same output, so this is a much simpler task than the one we are trying to accomplish here.} varying the value of the most relevant astrophysical parameters, discussed in Sec. \ref{sec:astro}: the threshold mass to host star formation in galaxies, $M_{min}$; the X-ray luminosity over star formation rate, $L_{X \leq 2 {\rm keV}}/{\rm SFR}$; and the number of ionizing photons per baryon, $N_{\gamma/b_\star}$. The value of the astrophysical parameters are laid down from a latin-hypercube with 1,000 elements that spans the range outlined in Table \ref{table:astro_params}. 

Since the effects arising by varying cosmological parameters are expected to be much smaller than the ones from astrophysics, we fix the value of those to the constraints from the Planck collaboration \citep{Aghanim:2018eyx}. Notice that, while the 21 cm field is highly affected by the different astrophysical processes, the matter density field is not, since \texttt{21cmFAST} does not account for backreaction of gas into dark matter.

For simplicity and computational efficiency, we work with 2D maps instead of 3D fields. For each simulation we take 20 2D slices orthogonal to the line-of-sight, in order to account for redshift-space distortions in the 21 cm field. Each slice contains $200\times200\times4$ voxels, that we project along the line-of-sight (i.e., summing the 4 voxels over the third dimension, where the redshift space distortions lie) to obtain 2D maps with $200\times200$ pixels. Each map represents thus a region of the Universe with dimensions $300\times300\times6~{\rm Mpc}^3$. We do that for each field in the simulation, namely the brightness temperature and the density field. We have verified that producing maps with slightly larger or smaller width along the line-of-sight does not change our conclusions.

Fig. \ref{fig:slices_redshift} shows an example of 21 cm and matter density fields at three different redshifts from one simulation. The effects of some of the cosmic processes introduced in Sec. \ref{sec:21cm} become patent. At $z \sim 20$, the 21 cm radiation is in the absorption regime, since the newborn galaxies start to emit UV radiation which redshifts towards Lyman frequencies and produces the Wouthuysen-Field effect. As new galaxies are formed, the UV radiation field increases and the Ly-$\alpha$ coupling becomes more important, reaching large amplitudes (in absolute value) at $z\sim 15$. At redshifts $z\sim 10$, the X-ray heating has become efficient enough to rise the temperature of the IGM and therefore reduce the absorption of the brightness temperature. Moreover, the ionizing radiation starts to spread abroad and expands the HII regions around the overdensities, appearing as the bright spots in the $\delta T_b$ panel. Note that this chronology strongly depends on the specific value of the astrophysical parameters; different values would lead to different timings of these evolutionary phases. On the other hand, the matter density field evolves quasi-linearly over time at high redshifts, entering into a non-linear regime at lower redshifts and on smaller scales.

\subsection{Network architecture}

\begin{table}
\setlength\extrarowheight{5pt}
\begin{center}
{
\begin{tabular}{|c||c|c|c| } 
 \hline
 \, Block \, & \, Layers \, \\
 \hline
 \hline
 \multirow{3}{*}{Unit layer}
 & Convolution layer \\
 & Batch normalization  \\
 & ReLU \\
 \hline
 \multirow{3}{*}{Contracting block}
 & Unit layer \\ 
 & Unit layer \\
 & Max-Pooling  \\
 \hline
  \multirow{4}{*}{Expanding block}
 & Unit layer \\ 
 & Unit layer \\
 & \; Transpose Convolution \; \\
 & Concatenation \\
 \hline
 \multirow{3}{*}{Final block}
 & Unit layer \\ 
 & Unit layer \\
 & Convolution layer \\
 \hline
\end{tabular}
}
\caption{\label{table:net}Different blocks used in the architecture of our neural network.}
\end{center}
\end{table}

We now outline the details of the model employed. As stated before, the goal is to undo the effect of astrophysics on 21 cm maps at redshifts $z\in[10-20]$, to recover the spatial distribution of the underlying matter field. We achieve this by training a deep convolutional neural network using the U-Net architecture \citep{2015arXiv150504597R}, firstly proposed in the context of biomedical image segmentation\footnote{Our implementation of the specific CNN architecture is based on the one in \url{https://github.com/Hsankesara/DeepResearch/tree/master/UNet}}. This kind of architecture has already applied to cosmological fields (see, e.g., \cite{He:2018ggn,Giusarma:2019feb}).
Fig. \ref{fig:architecture} shows an scheme of the network architecture, while each layer and block are detailed in Table \ref{table:net}. We define a \textit{unit layer} as the composition of a 2D convolutional layer, a batch normalization (BN) and a rectified linear unit (ReLU). Summarizing, the U-Net architecture consists of a contracting path, or encoder, followed by an expanding path, or decoder:
\begin{enumerate}
    \item The \textit{contracting path} is composed by the application of 3 contracting blocks, each of them composed by 2 unit layers followed by a 2x2 Max-Pooling operation. This reduces the spatial size by half.
    \item The \textit{expanding path} consists of the application of 3 expanding blocks, each of them composed by 2 unit layers followed by a transposed convolution layer for upsampling. The output of each upsampling is concatenated with the result from the contracting block (before pooling) of the same level, by means of \emph{skip connections}. Finally, there is a final block of 2 unit layers plus a final convolutional layer.
\end{enumerate}
For each convolutional layer, $3\times 3$ filters are employed, stride 1, applying a zero-padding with size 1. For the transpose convolutional layers, the filters also have a $3\times 3$ kernel, with stride 2 and padding 1. The network employed here has a total of 15x(Conv2D x ReLU x BN)+3 MaxPool + 3 transposed convolution = 51 layers and $\sim 8.5$ millions of parameters. Notice that the U-Net architecture is similar to the well-known Autoencoders (see, e.g., \cite{Goodfellow-et-al-2016}), with the main difference that here, layers between the encoder and the decoder are linked by means of the skip connections. This aspect is missing in typical Autoencoders, which are designed to recover an image with information only from the bottleneck. Notice also that in general, Autoencoders aim at producing the same output as the input, while this is not our purpose here.

\subsection{Training of the CNN}
\label{sec:training}

 As mentioned above, we make use of two different fields simulated by \texttt{21cmFAST}: 1) the 21 cm field $\delta T_b$ (the input) and the matter density $\delta$ (the output). We have employed maps from the $1,000$ numerical simulations to train, validate and test the network. 70\% of the maps are used for training, while 15\% for validation and the last 15\% for testing. We employ simulations at a fixed redshift for training. As stated before, we extract 20 2D slices from the 3D box, having therefore 20 slices per simulation. Moreover, we employ data augmentation to mock new data and to force the network to learn the symmetries of the problem, e.g. rotational and parity invariance.  We apply all the 8 possible rigid transformations in 2D to each slice: 4 rigid rotations with their respective reflections. Therefore, from each simulation box, we obtain $20 \times 8 =160$ slices for $\delta T_b$ and the same for $\delta$.

 We have trained our network throughout 40 epochs\footnote{We have verified that our results do not improve when training for more epochs.} with a batch size of 30, employing an Adam optimizer \citep{kingma2014method} with a learning rate of $10^{-3}$ and values $0.5$ and $0.999$ for the $\beta$ parameters\footnote{These are common hyperparameters, widely employed in deep learning literature. For definitions and detailed explanations, see, e.g., \cite{Goodfellow-et-al-2016}.}. Although the Adam optimizer already adapts the learning rate during the training, a scheduler has been applied in order to improve the convergence, reducing the learning rate by a factor of $10$ if the validation loss does not decrease after 10 epochs. For the loss function, we choose the mean squared error. We apply L2 regularization, with a weight decay value of $10^{-4}$. Our network and codes used for training, validation and testing are publicly available\footnote{\url{https://github.com/PabloVD/21cmDeepLearning}}.

We have explored the impact of variations in the architecture and hyperparameters on the final results. We have checked that, with enough epochs, the results are insensitive to remove the first concatenation of the U-Net. Similarly, a deeper network (i.e., including more unit layers, with corresponding pooling and concatenation links) do not produce any noticeable gain. Neither including more channels (i.e., enhancing the depth) nor using different activation functions, such as Leaky ReLU. Placing the batch normalization layer after the activation function does not have any impact in the results, giving equally good results.

\section{Results}

\label{sec:results}

\begin{figure*}[th]
    \centering
    \includegraphics[width=0.95\textwidth]{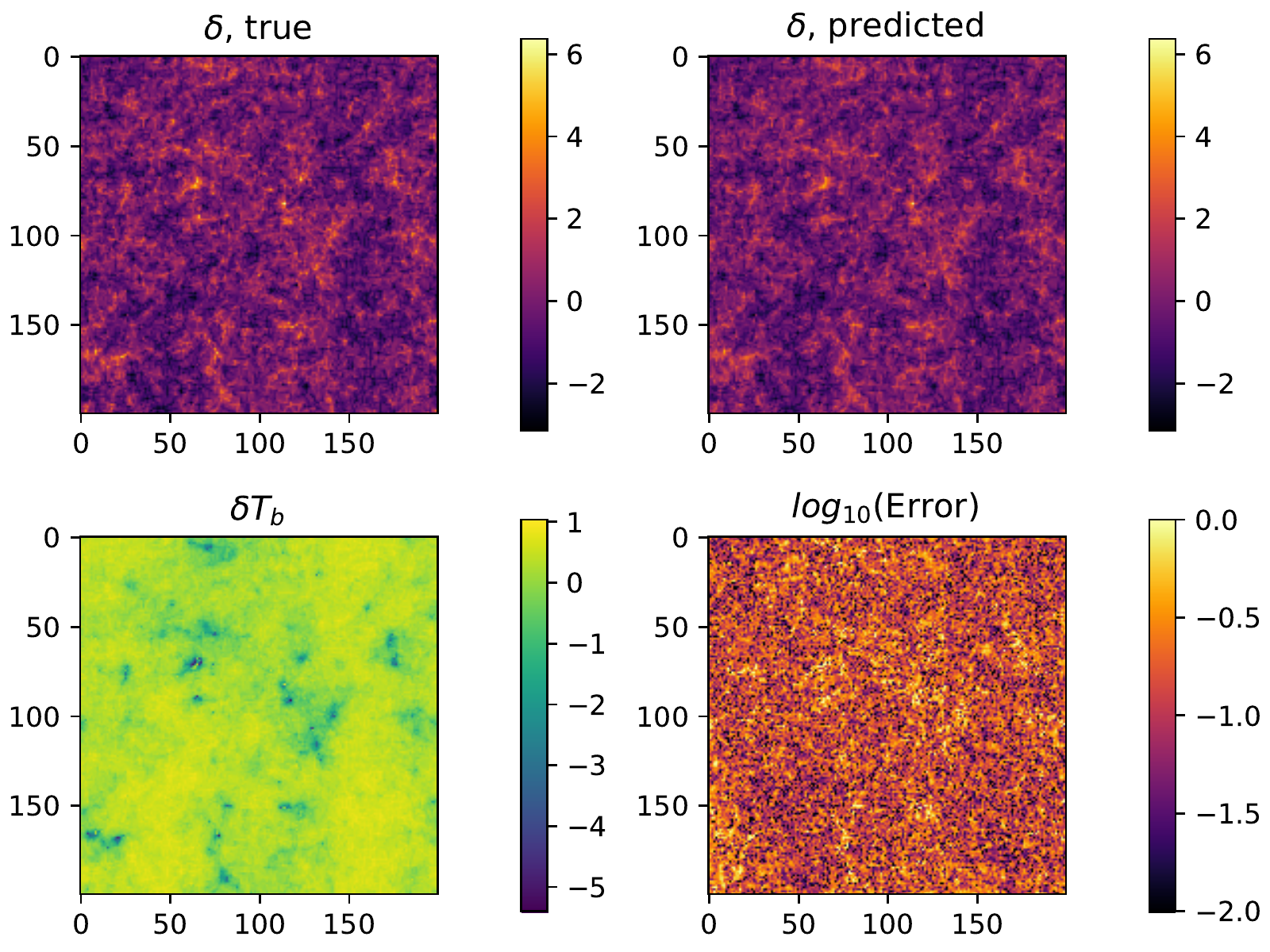}
    \caption{We train a neural network to output 2D matter density fields from 21 cm maps. The 21 cm maps have different values of the astrophysical parameters. Once the network is trained, we input a 21 cm map from the test set (bottom-left), which outputs the matter density field of the top-right panel. The true matter density map is shown in the top-left panel. Residuals between predicting and true is displayed in the bottom-right panel. As can be seen, the network produces matter fields that visually resemble very well the true ones. All results shown here are at $z=15$.}
    \label{fig:HI2DMmaps}
\end{figure*}

In this section we present the results of our analysis. Before quantifying the performance of the neural network, we can visually overview the success of the model at a sample of 2D slices from the same simulation in Fig. \ref{fig:HI2DMmaps}.

We train our network using pairs of 21 cm and matter density maps from the training set. Once the network has been trained, we input a 21 cm map from the test set (that the network has not seen before; bottom left panel) and the network produces an output (top right panel) that aims at matching the real underlying density field (top left panel). At a glance, the predicted field emulates with high accuracy the true field. This is confirmed looking at the logarithm of the absolute error $\log_{10}(\left\vert \delta_{\rm true}(\boldsymbol{r})-\delta_{\rm pred}(\boldsymbol{r}) \right\vert)$ (bottom right), which however takes larger values at the overdense spots in the $\delta$ maps, as one could naively expect.

In the following subsections we make use of different summary statistics to quantify the performance of the network. In the next subsections, we show results for a network trained with fields at $z\simeq 15$, except in subsections \ref{sec:z10} and \ref{sec:saliency}, where we compare the outputs at other times.

\subsection{Power spectrum and cross-correlation coefficient}

\begin{figure*}[th]
    \centering
    \includegraphics[width=\textwidth]{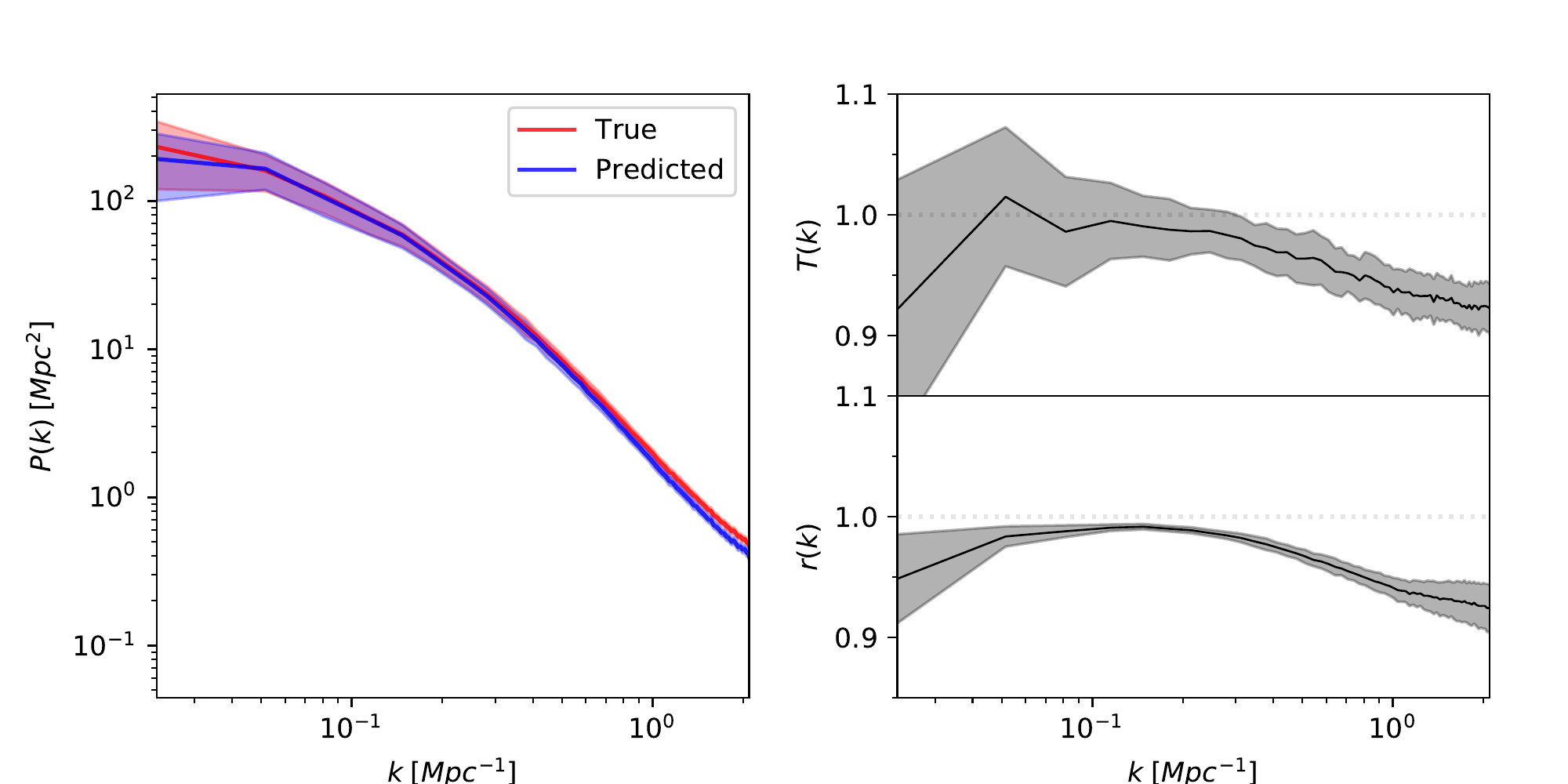}
    \caption{Once the network has been trained, we use it to produce matter density fields from 100 21 cm maps from the test set. The blue line in the left panel shows the mean of the power spectrum of these output maps, while its band represents the standard deviation. The red line and band display the results for the power spectrum of the true maps. The right panels show the transfer function (top-right), defined in Eq. \ref{Eq:Tk}, and the cross-correlation coefficient (bottom-right) defined in Eq. \ref{Eq:rk}. Our network is able to reconstruct the amplitudes and phases of the modes of the matter field within a few percent down to very small scales.}
    \label{fig:ps}
\end{figure*}

One of the most common used statistics in cosmology is the \textit{2-point correlation function}
\begin{equation}
    \xi(r) = \langle \delta(\boldsymbol{x}) \delta(\boldsymbol{x+r}) \rangle 
\end{equation}
or its Fourier transform, the \textit{power spectrum}
\begin{equation}
    P(k) = \int d^2r \; \xi(r) \; e^{i \boldsymbol{k \cdot r}}.
\end{equation}
Notice that in our definition we employ a differential element of area $d^2r$ instead of volume since we are working with 2D slices. We compute numerically the power spectra from our fields making use of the Python package \texttt{powerbox} \citep{G_Murray_2018}. In the left panel of Fig. \ref{fig:ps} we show the mean power spectrum over 100 samples of the testing dataset, with the bands representing the standard deviation among the testing slices. We can see that the power spectrum of the prediction from the CNN (blue line) reproduces with high accuracy the one of the real density field (red line). The prediction slightly worsens at small scales, presenting a 2$\sigma$ tension at $k \simeq 2$ Mpc$^{-1}$. To better quantify the deviation in the amplitude between both power spectra, we compute the \textit{transfer function} $T(k)$, defined as
\begin{equation}
    T(k) = \sqrt{ \frac{P_{\rm pred}(k)}{P_{\rm true}(k)} },
    \label{Eq:Tk}
\end{equation}
being $P_{\rm true}(k)$ and $P_{\rm pred}(k)$ the (auto-correlated) power spectra for the true and predicted fields respectively. The top-right panel of Fig. \ref{fig:ps} depicts this quantity as a function of scale, with its standard deviation region, showing that for all the scales, the transfer function is very close to 1, specially on the largest scales, where the perfect correlation case $T(k)=1$ is covered by the standard deviation region. Our model is able to recover the correct amplitude of the power spectrum within 5\% and 8\% down to $k\simeq0.7~{\rm Mpc}^{-1}$ and  $k\simeq2~{\rm Mpc}^{-1}$, respectively.

While the above transfer function informs us on the correlation between modes amplitudes, it is insensitive to modes phases. In order to quantify the accuracy on that, we made use of the cross-correlation coefficient, $r(k)$, defined as: 
\begin{equation}
    r(k) = \frac{P_{\rm true \times pred}(k)}{\sqrt{P_{\rm true}(k) P_{\rm pred}(k)}},
    \label{Eq:rk}
\end{equation}
being $P_{\rm true \times pred}(k)$ the cross-power spectrum between the true and the predicted fields. With this definition, a perfect match between the prediction and the true field would correspond to $r=1$. In bottom-right panel of Fig. \ref{fig:ps} we show $r(k)$ as a function of scale, which is also close to 1 for all scales, specially at $k \lesssim 3 \times 10^{-1}$ Mpc$^{-1}$, implying that there is a strong phase correlation between the prediction and the true fields. This can also be seen directly in Fig. \ref{fig:HI2DMmaps}. Similarly to the transfer function, our network matches the results from the true fields within 5\% and 8\% down to $k\simeq0.7~{\rm Mpc}^{-1}$ and  $k\simeq2~h{\rm Mpc}^{-1}$, respectively.

We note that the standard deviation of our results, both for the transfer function and the cross-correlation coefficient, increase on large scales. This reflects the effect of cosmic variance, that is larger on large scales. The deviations we find in both the transfer function and cross-correlation coefficient on small scales may be due to different effects. For instance, at small scales, non-linearities are more important, and the network could have more troubles mapping the density and the 21 cm field at that regime. Other reason may be that the effects of astrophysics strongly dominate on these scales, and they destroy the spatial correlations with the matter field. The different modifications of the network architecture commented at the end of Sec. \ref{sec:training} were not able to improve the performance. It would be desirable that a deeper and wider network, with a much finer tuning of the hyper-parameters, may improve our results. However, it may be that the effect of astrophysics smooths out correlations at small scales, and not significant improvement would be attainable.

\subsection{Probability Distribution Function}

\begin{figure}[th]
    \centering
    \includegraphics[width=0.45\textwidth]{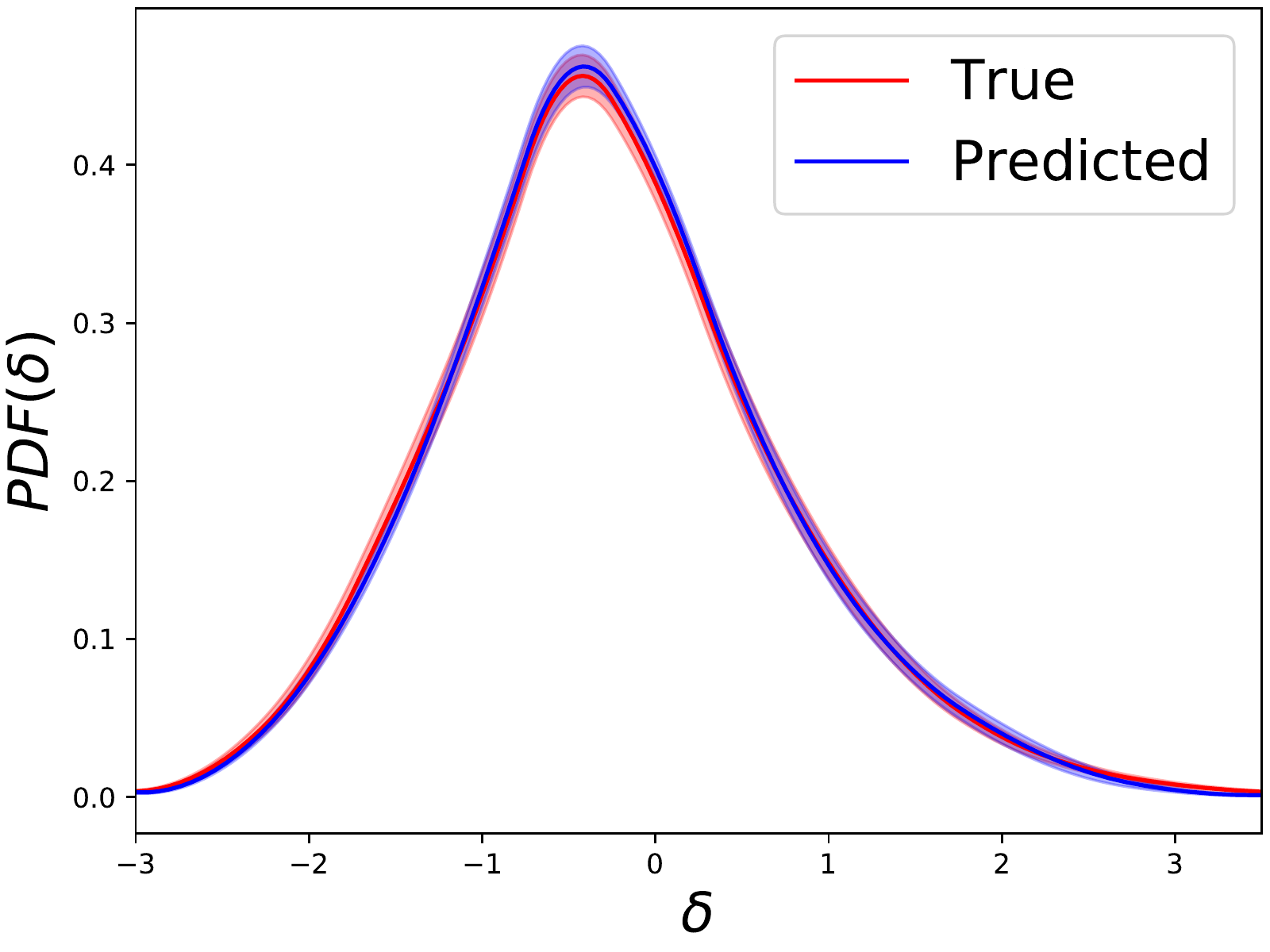}
    \caption{Probability Distribution Function (PDF) for the true (red) and predicted (blue) normalized density fields, with their corresponding standard deviation regions.}
    \label{fig:pdf}
\end{figure}

We now consider a different statistics that contains information complementary to the one of the power spectrum: the probability distribution function (PDF). Although the initial distribution which leave the seeds of the overdensities is taken as Gaussian centered around $\delta=0$, the evolution under non-linear dynamics modifies the shape of the PDF. Fig. \ref{fig:pdf} depict the PDFs for the true and predicted density fields, showing the resemblance between both cases. The PDF is computed in each sample image, being the solid lines and the shaded regions the mean and standard deviation respectively among the different testing images. We find a very good agreement between both PDFs, both in the peak and on the tails. To have a more quantitative insight of this similarity, the moments of both distributions have been computed, finding that the mean, variance and skewness of the PDFs coincide within the percent error, while the kurtosis presents a relative deviation of less than $\sim 10 \%$, with the predicted PDF presenting slightly less pronounced tails than the target one.

This shows that our network is able to capture the underlying distribution of the matter field, which at these redshifts start developing non-Gaussian tails.

\subsection{Astrophysical information embedded in the network}

\begin{table}
	\setlength\extrarowheight{5pt}
	\begin{center}
		\begin{tabular}{|c||c||c|}
			\hline
			Block & Output size & Weights\\ 
			\hline\hline
            Input & $1 \times 200 \times 200$ & -\\ 
            \; Encoder \; & \;  $256 \times 25 \times 25 \;$ & Fixed/Trainable \\
            \hline
            Unit layer & $ 512 \times 13 \times 13$ & Trainable \\
            Unit layer & $ 1028 \times 7 \times 7$ & Trainable \\
            Unit layer & $ 1028 \times 4 \times 4$ & Trainable \\
            Fully connected & $ 100$ & Trainable \\
            ReLU & $ 100$ & - \\
            Fully connected & $ 3$ & Trainable \\
			\hline
		\end{tabular}
	\end{center}
	\caption{\label{table:astronet} 
	Architecture of the convolutional neural network employed to predict the value of the astrophysical parameters from 21 cm maps. The Encoder is the contracting path of the U-Net shown in Fig. \ref{fig:architecture}.}
\end{table}

\begin{figure*}[th]
    \centering
    \includegraphics[width=\textwidth]{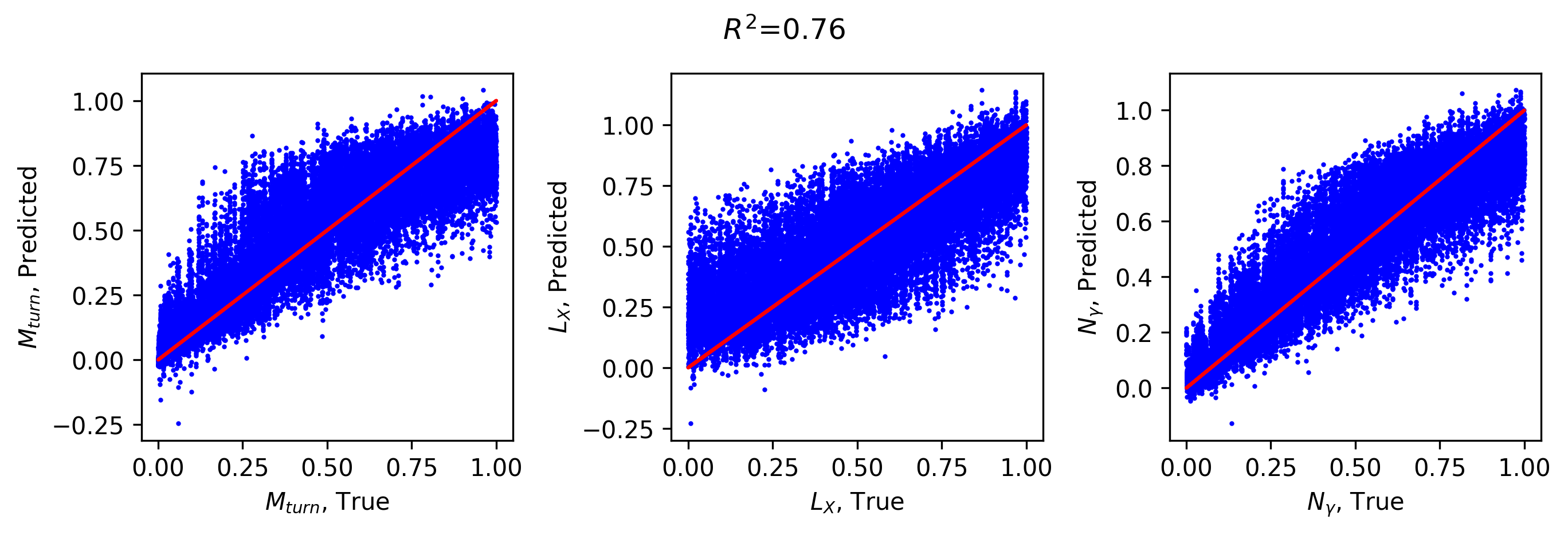}
    \includegraphics[width=\textwidth]{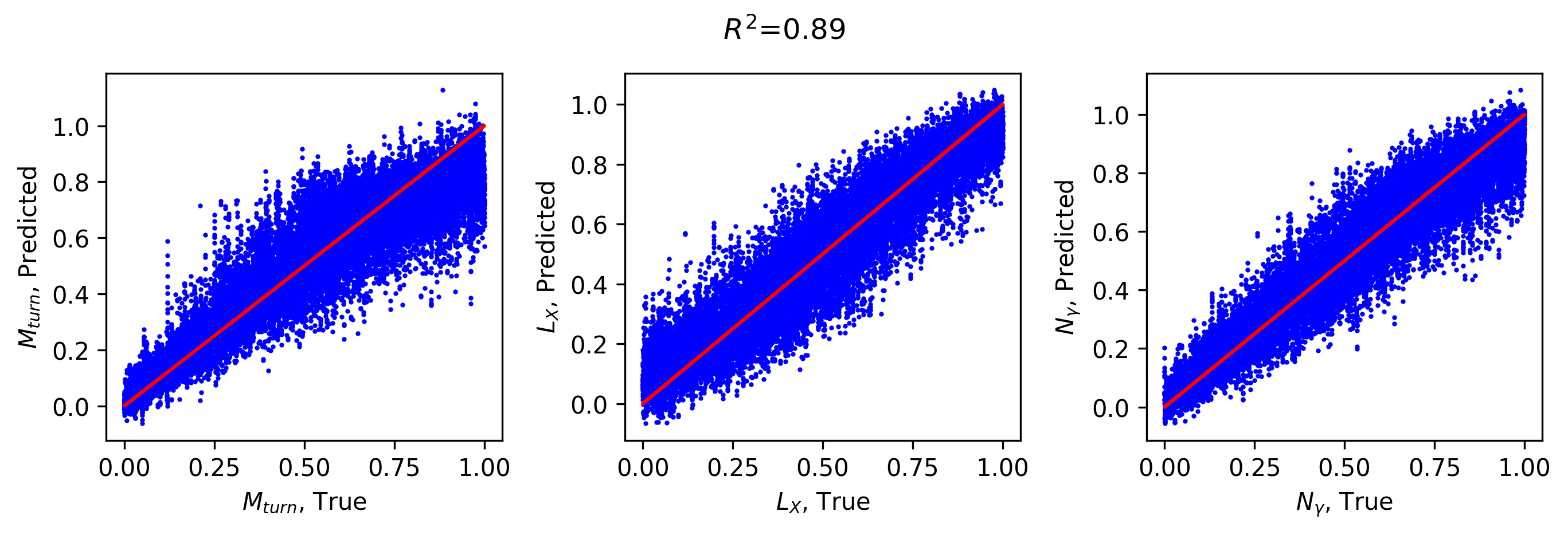}
    \caption{Our network is trained to output the matter field from 21 cm maps with different astrophysics. Thus, it is expected that the network will embed astrophysical information, that will use to undo its effects on 21 cm maps. We used a second neural network, that takes as input the output of the encoder part of the U-Net, and performs regression to the value of the astrophysical parameters. Results are shown in the top row. Notice that the weights of the encoder are fixed. We have repeated the same exercise but training also the weights of the encoder to perform regression; results are shown in the bottom row. We find that the encoder part of the U-Net indeed contains astrophysical information, although not as much as if the encoder were trained to perform parameter regression. Parameter values are normalized with respect to the mean. Results are shown at $z\simeq15$.}
    \label{fig:param}
\end{figure*}

We have shown in the previous sections that our network is able to learn the mapping between the 21 cm and density maps. To do so, it has to remove, or undo, the astrophysical processes. Thus, it is expected that different layers of the network carry out information about astrophysics, that the network will use to remove their effects. In order to check this, we have made use of an additional neural network to perform regression to the astrophysical parameters.

Specifically, we employ the result of the encoder, until the third Max-Pooling layer, to quantify the amount of information encoded in the contracting part of the U-Net. The procedure is as follows. First, a 21 cm map is input to the network. Next, we take the output of the encoder, and pass it to another neural network that predicts the value of the three astrophysical parameters of the input 21 cm map. While the encoder of the U-Net is fixed, i.e., its weights are kept fixed, the weights of the second network are tuned via gradient descent to match the value of the astrophysical parameters of the input 21 cm map.

This second network has the following architecture: 3 unit layers (i.e., 3 2D convolutional layers, each of them followed by a  Batch Normalization and a ReLu activation), employing stride 2 for the downsampling instead of pooling. The output is flattened and passed through a fully-connected layer with ReLU as activation function and dropout of $0.2$. Finally, a last fully-connected layer provides the 3 values for the astrophysical parameters. A scheme of the architecture is shown in Table \ref{table:astronet}. We employ the same loss function and optimizer than the used with the U-Net. We train for 20 epochs, and do not observe significant improvements for larger trainings.

The top panels of Fig. \ref{fig:param} show the predicted values of the astrophysical parameters versus their true values at $z\simeq 15$. 
We find that the network is able to approximate the proper mean value for each parameter, although with a large dispersion, having a linear correlation coefficient of $R^2 = 0.76$. We note this information is contained in the encoder arm of the U-Net, which has been previously trained to reproduce the density fields, i.e., the encoder part is not trained again with the loss function of the astrophysical parameters. 

The above experiment points out that our network indeed contains astrophysical information that can be used to regress the value of the astrophysical parameters. However, the network is not trained to achieve that, so it is expected that a network specifically trained to perform parameter regression may perform better. In order to verify that, we have repeated the above experiments but training also the weights of the U-Net encoder.

The results of such tests are shown in the bottom row of Fig. \ref{fig:param}, where the linear correlation coefficient is enhanced up to $R^2 = 0.89$, and the accuracy of the network has improved significantly. This suggests that the encoder trained to predict the matter density field does not carry all the possible astrophysical information. We note that more astrophysical information may be embedded into the decoder part, that we are not quantifying here. However, it may be that, rather than learning directly the astrophysical information, the U-Net is able to reproduce the density field by other  means.

We have verified that different variations on the architecture, such as adding or removing convolutional, pooling or fully connected layers, do not significantly improve our results.

Notice that previous works have already employed neural networks to predict the underlying astrophysical models \citep{Schmit:2017pho,Kern:2017ccn,Shimabukuro:2017jdh,Gillet:2018fgb,Hassan:2018uhw,Hassan:2019cal}, obtaining much better accuracies than the ones shown here. However, these works employ inputs at several redshifts, as opposed to our case, since we want to explore the amount of astrophysical information encoded in the U-Net. Nonetheless, as we show in the next section, we can recover the astrophysical parameters with high accuracy at lower redshifts.

\subsection{Redshift dependence}
\label{sec:z10}

\begin{figure*}[th]
    \centering
    \includegraphics[width=\textwidth]{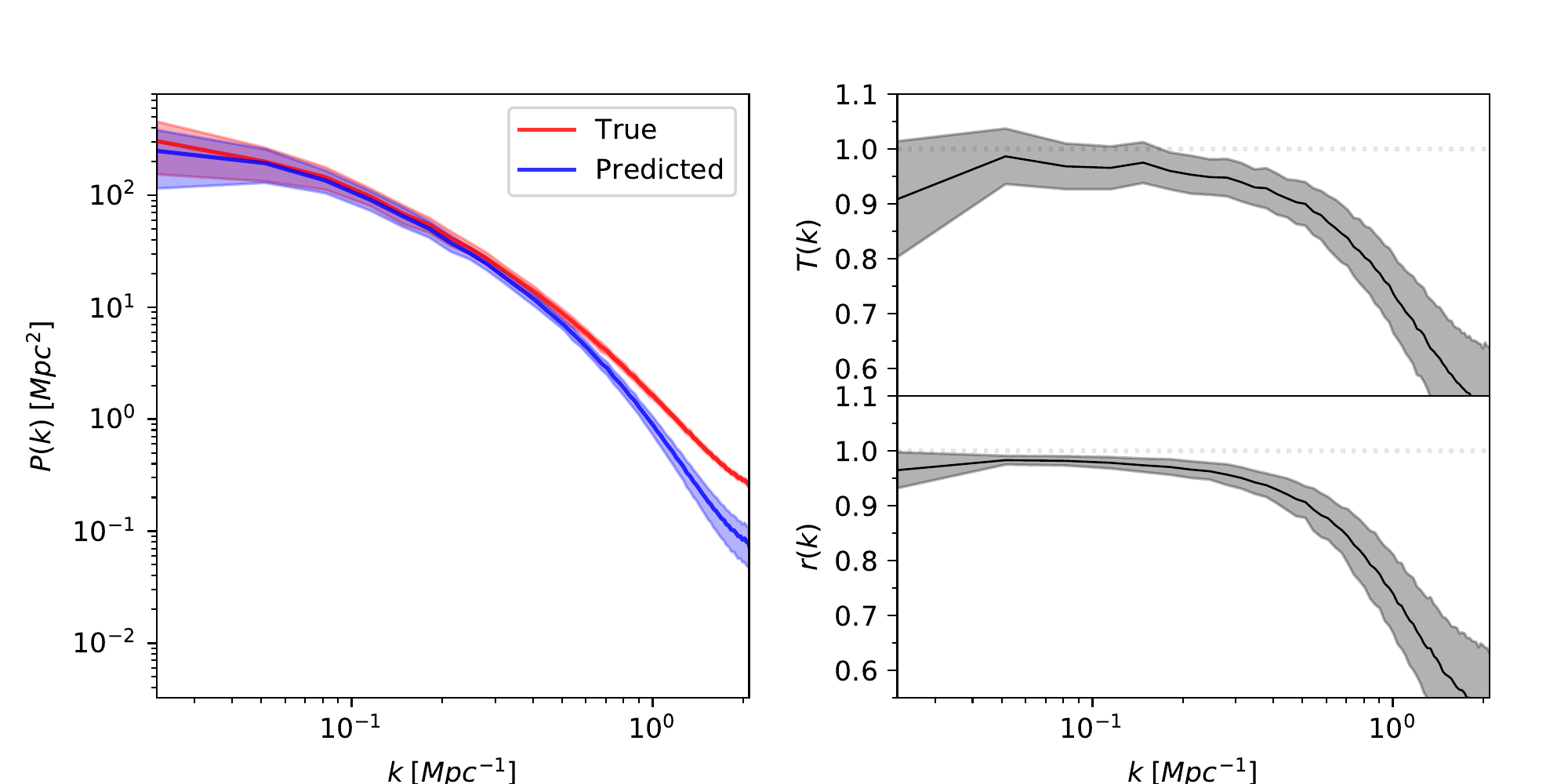}
    \caption{Same than Fig. \ref{fig:ps} but at $z\simeq 10$.}
    \label{fig:ps_10}
\end{figure*}

\begin{figure*}[th]
    \centering
    \includegraphics[width=\textwidth]{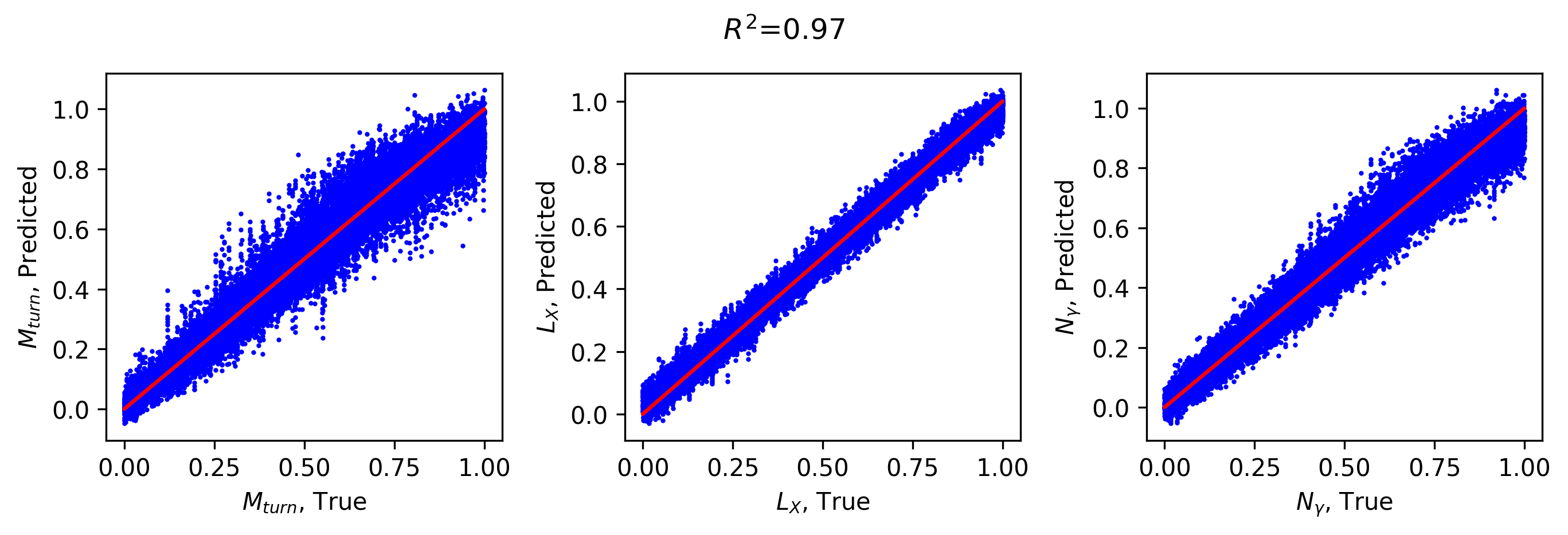}
    \caption{Same than bottom panel of Fig. \ref{fig:param} but at $z\simeq 10$.}
    \label{fig:param_10}
\end{figure*}

In the previous subsections we have focused our analysis at $z \simeq 15$. Here we investigate the dependence on redshift of our results.

We have checked that at higher redshifts, $z \simeq 20$, we still recover the different statistics considered with a similar accuracy than at $z\simeq15$. However, at lower redshifts, the performance of the U-Net worsen significantly. To illustrate this, we show results for the power spectrum, transfer function and cross-correlation coefficient at $z\simeq 10$ in Fig. \ref{fig:ps_10}. Although the amplitude of the power spectrum is still well recovered on large scales, it is further suppressed for wavenumbers lower than $\simeq3\times 10^{-1}$ Mpc$^{-1}$. Both $T(k)$ and $r(k)$ depart significantly from $1$ on small scales.

We believe that the reason of the worse performance is mainly driven by the presence of Reionization processes. At this low redshifts, the release of UV radiation from galactic sources is efficient enough to ionize a significant portion of the Universe, forming HII bubbles around each source. These bubbles start to grow and overlap at this epoch, although they may not be completely ionized yet. In an analogous way, temperature inhomogeneities driven by the heating processes may become more important at these stages too, which contribute to the 21 cm maps. Therefore, the link between the differential brightness temperature and the matter density field gets weaker and $\delta$ cannot be recovered as accurately as in previous times, when the astrophysical processes were not so advanced.

However, while it is more difficult to reproduce the matter density field from the brightness temperature, we find that the astrophysical parameters can be predicted with much higher accuracy, as we show in Fig. \ref{fig:param_10} for maps at $z\simeq 10$. For this regression, we re-train the same neural network of the previous section, training at the same time the U-Net encoder and the final layers, as was showed to achieve better results. 

The linear correlation coefficient reaches now $R^2=0.97$. This better behaviour can be explained for the same reason we find worse results in the U-Net. The astrophysical processes are more advanced at these stages of the cosmic history. Reionization and heating processes leave significant imprints in the brightness temperature field, which are now more patent, and strongly dependent on the astrophysical model. This allows the network performing regression to identify the underlying model more easily, while is harder to find correlations with the underlying matter field. It is worth to note that this accuracy has been achieved employing only one redshift, being comparable or even better to other cases in the literature (e.g., \cite{Gillet:2018fgb}) where inputs at several redshifts have been used. The above results suggest the trend at lower redshifts and scales that the 21 cm signal keeps astrophysical information to the detriment of the density field. It remains as an open question whether there is a limiting scale up to which the maximum information achievable can be extracted, regardless of the training and tuning of the neural network.

\subsection{Saliency maps}
\label{sec:saliency}

\begin{figure*}[th]
    \centering
    \includegraphics[width=0.32\textwidth]{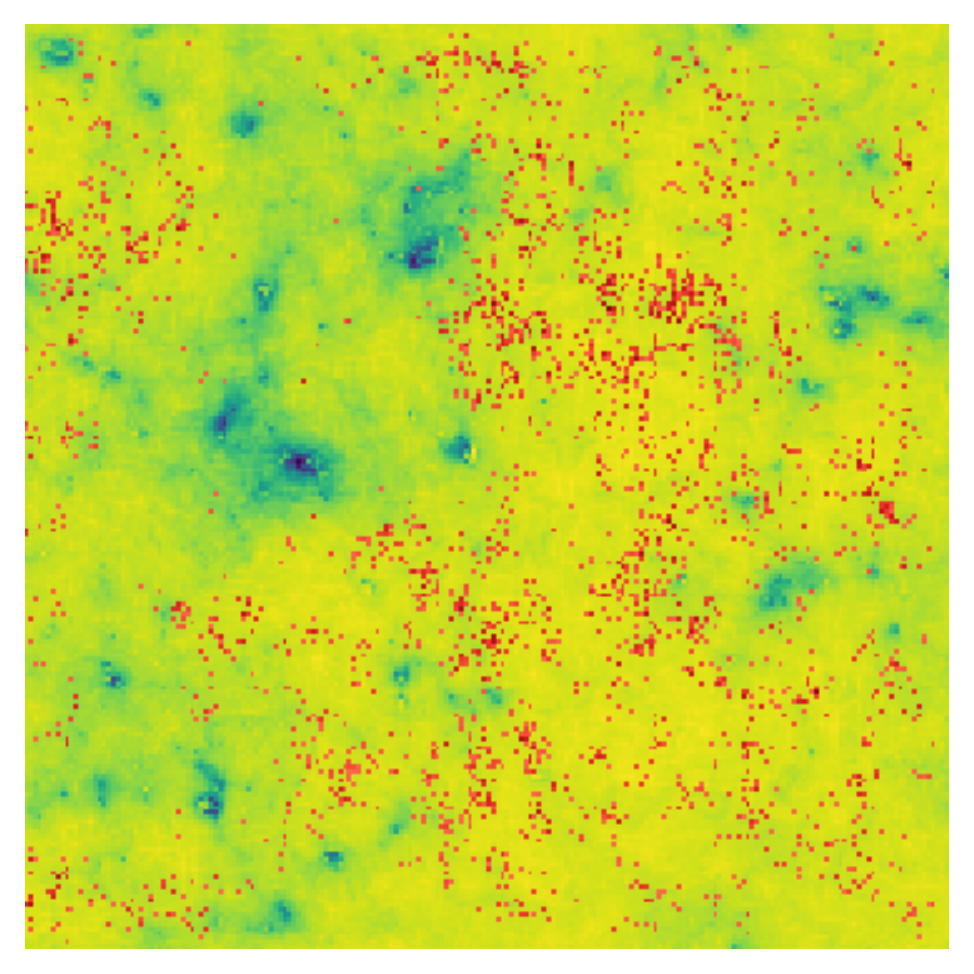}
    \includegraphics[width=0.32\textwidth]{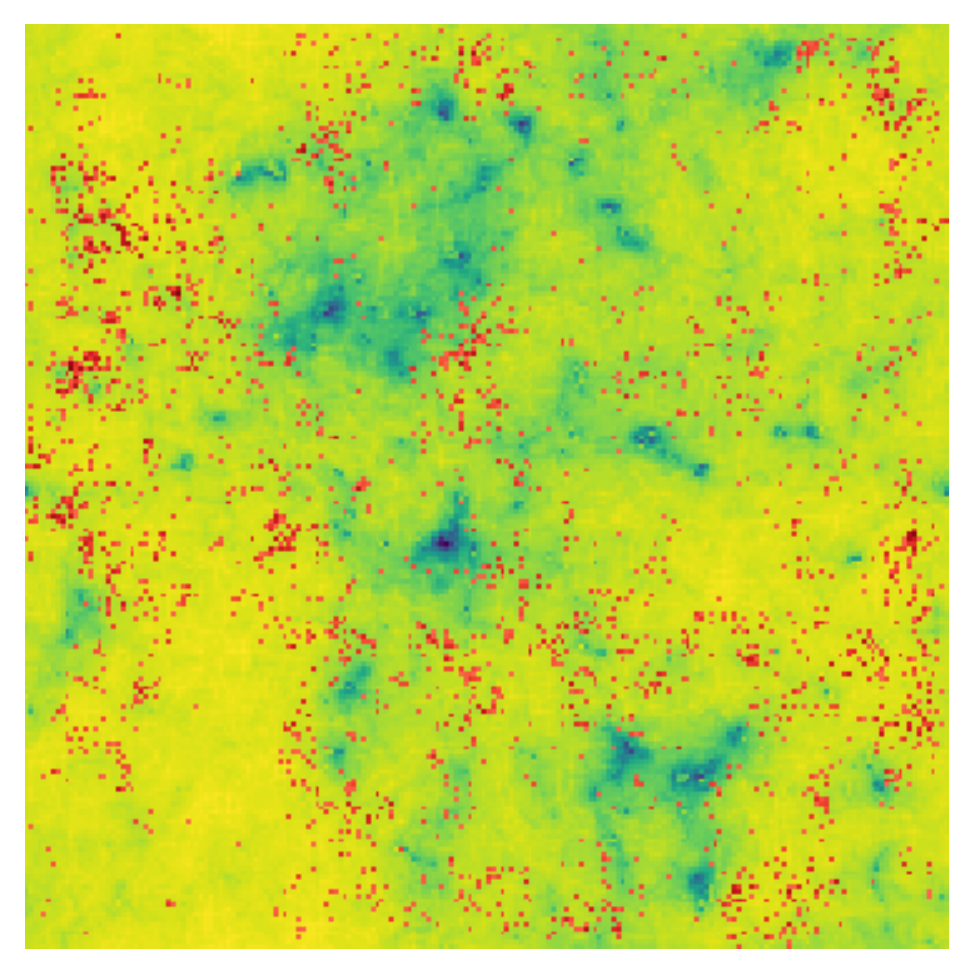}
    \includegraphics[width=0.32\textwidth]{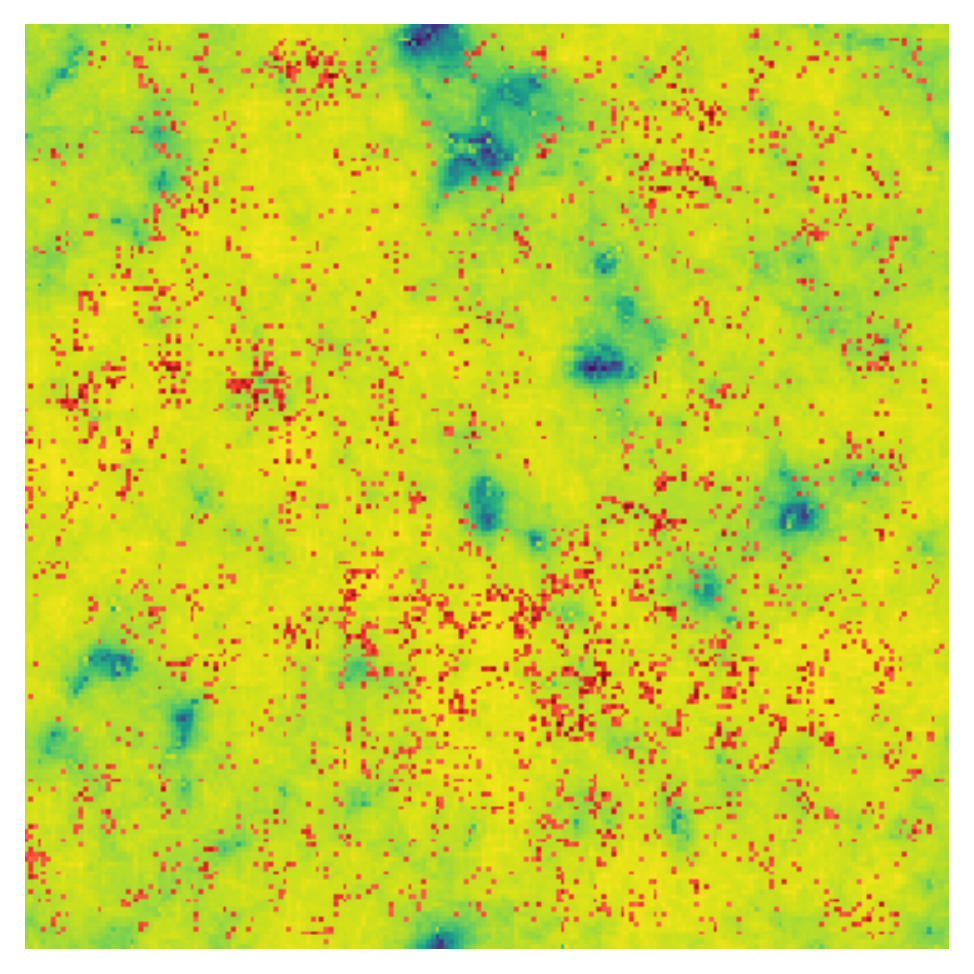}
    \includegraphics[width=0.32\textwidth]{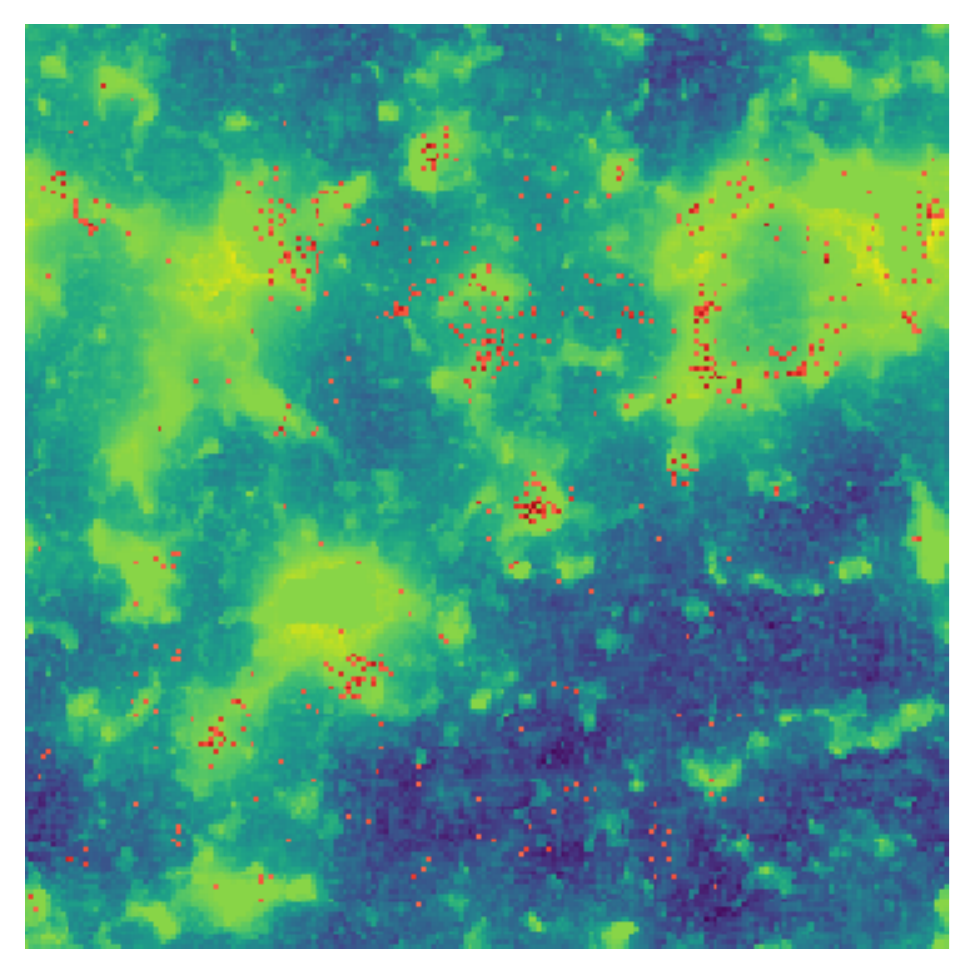}
    \includegraphics[width=0.32\textwidth]{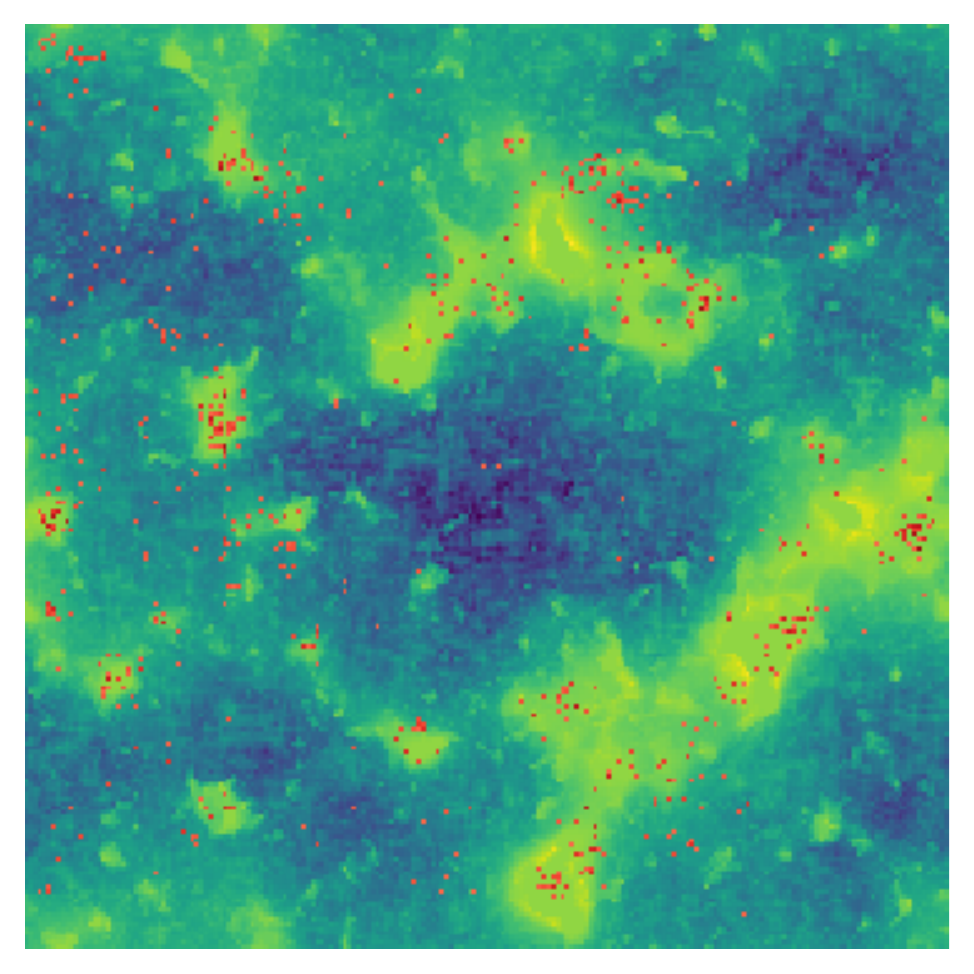}
    \includegraphics[width=0.32\textwidth]{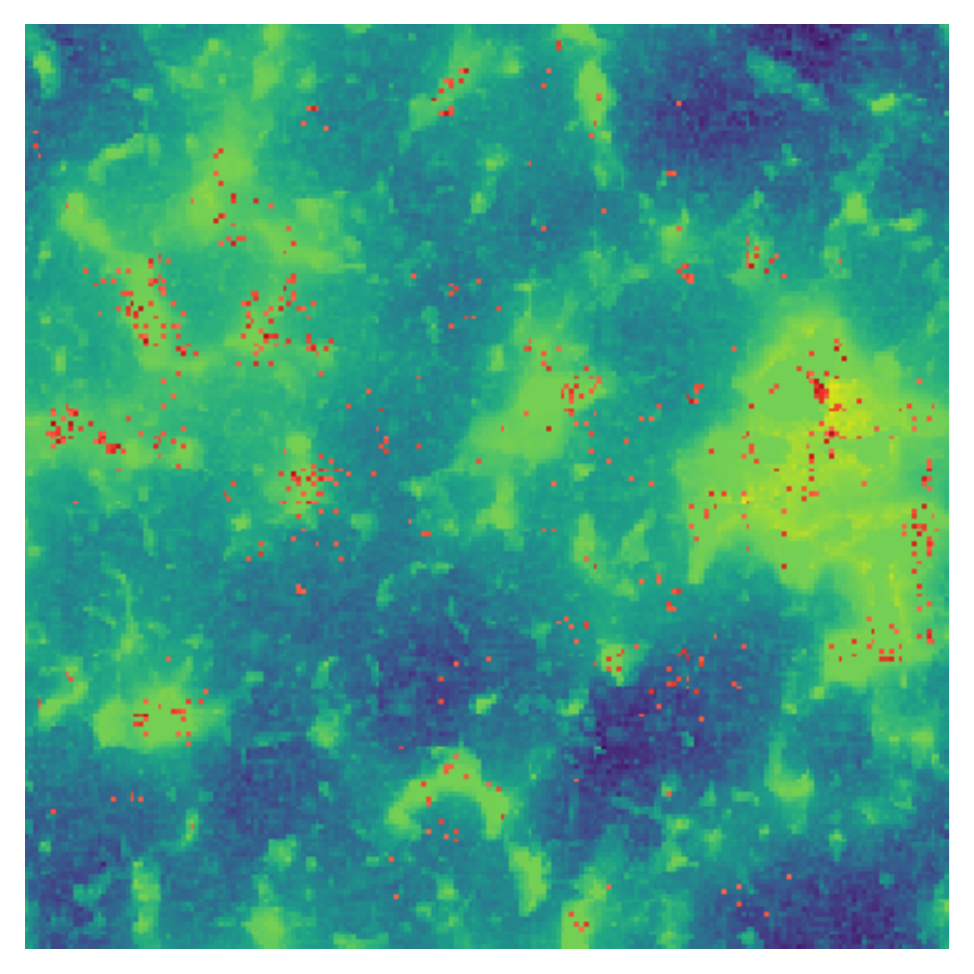}
    \caption{Samples of 21 cm brightness temperature $\delta T_b$ maps at $z\simeq 15$ (top) and $z\simeq 10$ (bottom). The brightest pixels of their respective saliency maps for $M_{\rm turn}$ are superimposed as red pixels. The saliency maps are computed from the gradient of the outputs (the astrophysical parameters) respect to the input image. Brighter pixels in the saliency map denote regions which have more impact on the output of the neural network. The maps for the other astrophysical parameters are very similar at $z\simeq 15$, although they differ at lower redshifts, as can be seen in Fig. \ref{fig:saliency_3params_z10}. While at $z\simeq15$ the network seems to be looking at regions away from the sources, at lower redshifts the network appears to focus its attention into the heated and ionized regions.}
    \label{fig:saliency}
\end{figure*}

Although neural networks have been shown to be very successful at finding correlations and extracting information from the data, understanding what operations or features they are looking up is not an easy task. Saliency maps provide a way to quantify what pixels in the original image are the ones that have the largest influence on the output of the network \citep{simonyan2013deep}. Here, we employ this method with the neural network devoted to predict the value of the astrophysical parameters, in order to try to understand what the network is looking at. Since training the encoder together with the top layers of the net provided better results, we employ that version of the astrophysical network for this test.

The saliency maps are generated as follows. First, with the neural network trained, we input a 21 cm map. Next, 3 saliency maps are generated by computing the derivative of the output, with respect to the input. In our case, the output is just the value of the astrophysical parameters, while our input is the 21 cm brightness temperature in each pixel of the image.

The pixels whose gradient is large (in modulus) are the ones whose variations will most largely impact the output of the network. For example, in typical classification problems in computer vision, this procedure usually tends to produce larger gradients for pixels where the object to be classified is located. The algorithm that we follow for producing the saliency maps is the so-called Vanilla Gradient. Although there are further improvements on this basic idea, Vanilla Gradient has been shown to be robust and fast, also avoiding some problems that other techniques may present \citep{DBLP:journals/corr/abs-1810-03292}.

Fig. \ref{fig:saliency} shows samples of 21 cm maps from our simulations. The red dots in the images show the locations of the pixels with the largest absolute values of the saliency maps for $M_{\rm turn}$. Darker points indicate larger gradients in modulus. Top panels show results at $z\simeq 15$, while bottom panels correspond to $z \simeq 10$. We find that the output of the saliency maps are redshift dependent. At $z\simeq 15$, we can clearly see in top panel of Fig. \ref{fig:saliency} that the brightest pixels of the saliency map (red points) correspond to regions in the 21 cm field relatively far from the overdensities, presenting therefore less absorption. Instead of looking at the radiative sources, as one could naively expect, the neural network seems to extract most of the astrophysical information from the underdense regions around them. On the other hand, at lower redshifts $z\simeq 10$ (bottom panel of Fig. \ref{fig:saliency}), when the astrophysical processes are more advanced, the neural network pays more attention to regions, or the edges of these regions, which are more heated and ionized, where the astrophysical processes are leaving stronger signatures. Nonetheless, in both cases the saliency maps highlight neutral regions far away from the sources. We note that further work is needed in order to extract more robust conclusions on where the network information is coming from.

\begin{figure*}[th]
    \centering
    \includegraphics[width=\textwidth]{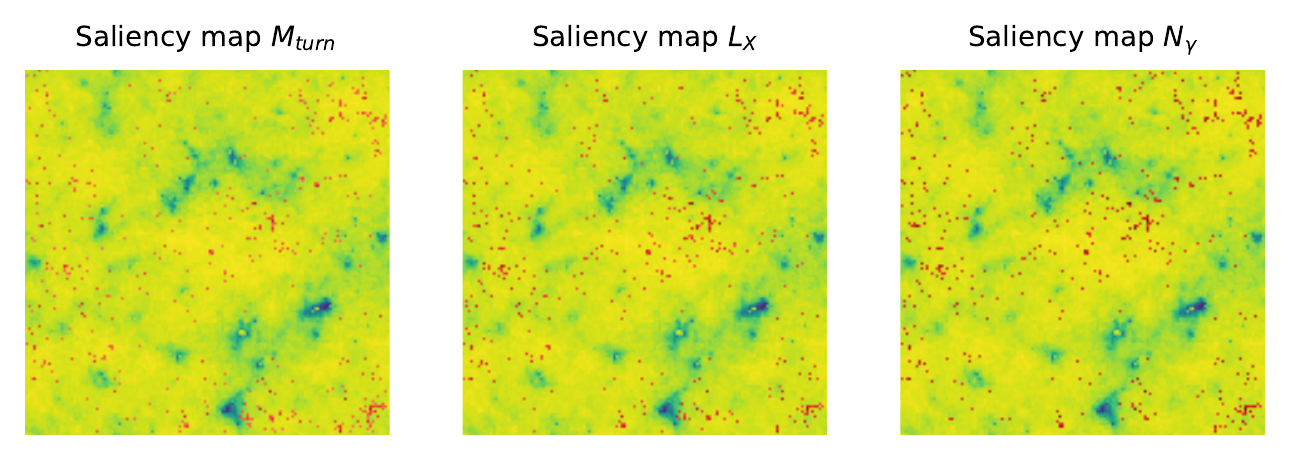}
    \includegraphics[width=\textwidth]{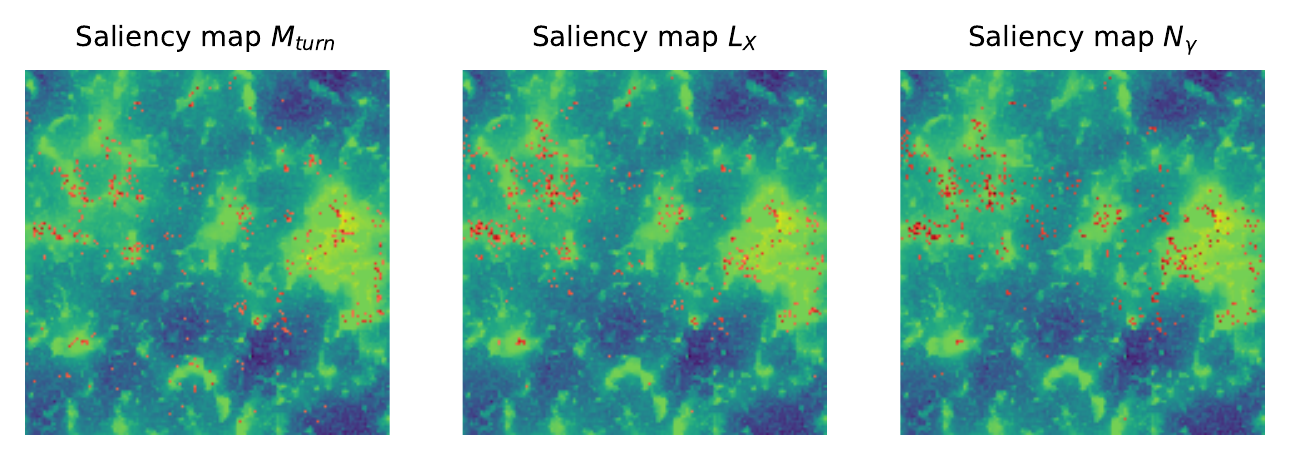}
    \caption{21 cm maps, together with the brightest pixels of their corresponding saliency maps (red dots), for each astrophysical parameters at $z\simeq 15$ (top) and at $z\simeq 10$ (bottom). The network seems to be looking at the same regions when performing regression to the three different astrophysical parameters.}
    \label{fig:saliency_3params_z10}
\end{figure*}

The above results stand only for the parameter $M_{\rm turn}$, but in principle, the saliency maps for the other parameters may present differences among them, since the different astrophysical processes may induce distinct features on the brightness temperature. However, we find the saliency maps for the three parameters to be pretty similar (Fig. \ref{fig:saliency_3params_z10}), which indicates that the neural network employs the same pixels of the input image for predicting any of the parameters.  Nonetheless, the magnitude of the gradients varies among the different parameters. As can be seen in Fig. \ref{fig:saliency_3params_z10}, the points in the $N_{\gamma}$ saliency map are redder and more clustered, meaning that this parameter could be more sensitive to variations of the pixels in the input image.

All these results present a step forward towards the search for an optimal estimator to extract astrophysical information from 21 cm maps. However,  further research is needed in order to get a better understanding on that estimator and its properties.

\section{Conclusions}
\label{sec:conclusions}

Brightness temperature fluctuations in 21 cm maps contain a rich amount of information on both cosmology and astrophysics. Unfortunately, they are coupled in a non-trivial way. In this paper we have used deep convolutional neural networks to recover 2D images of the underlying matter field (in real-space) from 21 cm maps in redshift-space. Being able to generate 2D matter density fields at high-redshifts will be beneficial for several reasons: 1) the power spectrum will allow to extract most of the cosmological information down to pretty small scales, 2) they can be used to improve our knowledge on the halo-galaxy connection, and 3) they will allow us to better understand the impact of astrophysics on 21 cm maps.

We have run a set of 1,000 numerical simulations, varying the value of 3 astrophysical parameters in a wide range. The value of those parameters are laid down in a latin-hypercube. From each simulation we generate both 21 cm maps and their corresponding matter density fields.

We have trained U-Net architecture to find the mapping between 21 cm maps and 2D matter density fields. Since each 21 cm map is affected by astrophysics in a different manner, the network is forced to undo astrophysical effects.

We find that at $z=15$, our network is able to generate 2D density maps whose power spectra agree with the true ones at $\simeq8\%$ down to $k\simeq2~{\rm Mpc}^{-1}$, with a much better performance on larger scales. Similar results hold for the cross-correlation coefficient, showing that the network is able to reconstruct both modes amplitudes and phases with high-accuracy. Other statistics of the generated maps, e.g. the 1-point PDF are also in excellent agreement with those of the true maps.

At lower redshifts our results become worse, likely due to the effect of non-linearities in the matter field, but mainly because astrophysics effects become so large that the spatial correlations between the 21 cm and matter field are largely affected and become weaker.

It is expected that different layers of our network carry out information on astrophysics, as that is needed to undo their effects. We have verified this by taken the output of the U-Net encoder part and perform regression to the value of the astrophysical parameters through a secondary neural network. We find that this set up allow to place constraints on the value of the astrophysical parameters, pointing out the presence of astrophysical information in the network.

That information is however not the maximum available. We have tested this by retraining the above model, but not keeping fixed the value of the encoder weights (i.e., training the encoder weights at the same time as the second network). In that case we are able to get more accurate constraints on the parameters. This indicates that while the network carry out astrophysical information, the encoder part does not maximize it for the regression task. The decoder part may contain additional astrophysical information, or the network does not need to maximize regression information to perform the mapping to the matter field.

Finally, we have made use of saliency maps to investigate the features and pixels the network is giving more weight when performing regression to the parameters. For simplicity, we use the network where the encoder is trained at the same time as the secondary network. At redshfits $z\simeq15$, we find that the network is most sensitive to the 21 cm pixels that are far away from sources, while at lower redshfits, the network seems to be focusing most of its attention into the heated and ionized regions around the sources (albeit not completely ionized yet). The usage of saliency maps, or other tools developed to interpret neural networks behaviour, is an important step towards identifying the best statistics to extract information from 21 cm maps.

It is important to emphasize the simplifying assumptions we have made in this work. First of all, our 21 cm maps do not include any instrumental noise. In real observations, this noise will significantly affect the temperature fluctuations of the 21 cm maps. Residual foreground removal is also expected to affect the 21 cm signal on large scales. However, these effects would mainly impact the power spectrum at small scales, where the network is not as reliable. One may deal with this issue imposing a cut at the scales where we trust the recovered statistics. Nevertheless, a complete analysis including noises and foreground would be mandatory in order to apply this network to cosmological observations.

 We have used \texttt{21cmFAST} to generate the density and 21 cm fields, with a minimal choice of three free astrophysical parameters. Full hydrodynamic simulation may produce slightly different results with more complex patterns for the 21 cm field and its relation with the underlying matter field. Furthermore, the density fluctuations are computed using second-order lagrangian perturbation theory, which suppress power at small scales with respect to full N-body simulations. This may affect the performance of the network at these scales when more accurate density simulations are employed. Finally, while in this work we have changed the value of the astrophysical parameters within a wide range, we have kept fixed the cosmology. Whereas it is expected that changes in cosmology within current bounds will have a much smaller effect than astrophysics, it should be explicitly tested how much this affects our results.

All the above effects will degrade the accuracy with which we can recover the underlying matter density field. On the other hand, we have performed our analysis on 2D, where some information is loss due to projection effects. A neural network trained to find the mapping between the 3D 21 cm and matter fields may improve our results. Furthermore, our neural network has trained using maps at a single redshift; using maps, or 3D fields, at different redshifts may improve the overall performance of the network, in particular at lower redshifts. We leave for future work a quantification of all these different issues.

\section*{Data availability}
The codes and networks underlying this article are available in \url{https://github.com/PabloVD/21cmDeepLearning}.

\section*{Acknowledgements}  
PVD thanks the Department of Astrophysical Sciences of Princeton University for its hospitality during the early stages of this work. This work has made use of the Tiger cluster of Princeton University. We thank Gabriella Contardo, Yin Li, Shirley Ho, and David Spergel for useful conversations. FVN acknowledge funding from the WFIRST program through NNG26PJ30C and  NNN12AA01C. PVD is supported by the Spanish MINECO grants SEV-2014-0398 and FPA2017-85985-P, and by PROMETEO/2019/083.

\bibliography{biblio}

\end{document}